\documentclass[12pt]{article}

\begin{document}

\title{\bf \Large Non-Static Spherically Symmetric Perfect Fluid
Solutions}

\author{M. Sharif \thanks{e-mail: hasharif@yahoo.com} and T.
Iqbal
\\ Department of Mathematics, University of the Punjab,\\ Quaid-e-Azam
Campus Lahore-54590, PAKISTAN.}

\date{}

\maketitle

\begin{abstract}
We investigate solutions of Einstein field equations for the non-static
spherically symmetric perfect fluid case using different equations of
state. The properties of an exact spherically symmetric perfect fluid
solutions are obtained which contain shear. We obtain three different
solutions out of these one turns out to be an incoherent dust solution
and the other two are stiff matter solutions.

\end{abstract}

\section{\textbf{INTRODUCTION}}

There is no shortage of exact solutions of Einstein field equation (EFEs).
However, because General Relativity is highly non-linear, it is not
always easy to understand what qualitative features solutions might possess.
Different people have been working on the investigation of spherically
symmetric perfect fluid solutions with shear [1-5]. Nearly all the
solutions have been obtained by imposing symmetry conditions. It is known that
non-linear partial differential equations admit large classes of solutions,
many of which are unphysical.

EFEs for static spherically symmetric distribution of perfect fluid have been investigated by many
authors using different approaches [6]. One approach is to prescribe an equation of state, $\rho = \rho
(p)$ which relates the energy density $\rho$ and isotropic pressure p. In this paper we shall extend
this idea of solving EFEs for non-static spherically symmetric spacetimes. We shall examine
systematically the field equations for non-static spherically symmetric
perfect fluid solutions. We obtain three different solutions which have shear non-zero. Since most of
the models in the literature [6] for spherically symmetric are shear-free. Thus it will be interesting
to study the non-static solutions which contain shear.

The breakup of the paper is as follows. In section two we shall write down
the field equations. In the third section we attempt all the possible
solutions in three classes using different equations of state. In section
four we shall evaluate kinematic quantities for the solutions obtained.
Finally in the last section we shall conclude our discussion.

\section{\textbf{Field Equations}}

The non-static spherically symmetric metric has the form [5]

\begin{equation}
ds^2=e^{2\nu (r,t)}dt^2-e^{2\lambda (r,t)}dr^2-R^2(r,t)d\Omega ^2,
\end{equation}
where $d\Omega ^2=d\theta ^2+\sin ^2\theta \,d\varphi ^2$.

For a perfect fluid distribution, the energy-momentum tensor is
given by
\begin{equation}
T_{ab}=(\rho+p)u_a u_b -pg_{ab},\quad u_a u^b=1, \quad a,b=0,1,2,3.
\end{equation}
where $\rho$ and $p$ are the energy density and the pressure of the fluid
respectively. The four velocity of the fluid has the form $u^a
=(e^{-\nu(r,t)},0,0,0)$. The field equations for the metric (1) can be written down [6]
\begin{equation}
\kappa \rho =\frac 1{R^2}-\frac 2Re^{-2\lambda }(R^{^{\prime \prime
}}-R^{^{\prime }}\lambda ^{^{\prime }}+\frac{R^{^{\prime }2}}{2R})+\frac
2Re^{-2\nu }(\stackrel{.}{R}\stackrel{.}{\lambda }+\frac{\stackrel{.}{R}^2}{%
2R}),
\end{equation}
\begin{equation}
\kappa p=-\frac 1{R^2}+\frac 2Re^{-2\lambda }(R^{^{\prime }}\nu ^{^{\prime
}}+\frac{R^{^{\prime }2}}{2R})-\frac 2Re^{-2\nu }(\stackrel{..}{R}-\stackrel{%
.}{R}\stackrel{.}{\nu }+\frac{\stackrel{.}{R}^2}{2R}),
\end{equation}
\[
\kappa pR=e^{-2\lambda }((\nu ^{^{\prime \prime }}+\nu ^{^{\prime }2}-\nu
^{\prime }\lambda ^{^{\prime }})R+R^{^{\prime \prime }}+R^{^{\prime }}\nu
^{^{\prime }}-R^{^{\prime }}\lambda ^{^{\prime }})
\]
\begin{equation}
-e^{-2\nu }((\stackrel{..}{\lambda }+\stackrel{.}{\lambda }^2-\stackrel{.}{%
\lambda }\stackrel{.}{\nu })R+\stackrel{..}{R}+\stackrel{.}{R}\stackrel{.}{%
\lambda }-\stackrel{.}{R}\stackrel{.}{\nu }),
\end{equation}

\begin{equation}
\stackrel{.}{R}^{^{\prime }}-\,\stackrel{.}{R}\nu ^{^{\prime }}-\stackrel{}{%
R^{^{\prime }}}\stackrel{.}{\lambda }=0,
\end{equation}
where the dot denotes partial derivative with respect to time `$t$' and the
prime indicates partial derivative with respect to the coordinate `$r$'. The
spatial coordinate `$r$' refers to the comoving radius and `$\kappa $' is the
gravitational constant. The consequences of the energy-momentum
conservation $T_{;b}^{ab}=0$ are the relations
\begin{equation}
p^{^{\prime }}=-(\rho +p)\nu ^{^{\prime }},\quad\stackrel{.}{\rho }%
=-(\rho +p)(\stackrel{.}{\lambda }+2\frac{\stackrel{.}{R}}R).
\end{equation}

We now consider the equation of state
\begin{equation}
p=(\gamma -1)\rho ,\quad\rho +p=\rho \gamma , \quad 1\leq \gamma \leq 2,
\end{equation}
where $\gamma$ is a constant. (The limits on $\gamma$ result from the
requirement that the stresses be pressures rather than tensions and
that the speed of sound in the fluid be less than the speed of light
in vacuum). For $\gamma =1$, the pressure vanishes, so that the
equation of state is that of incoherent dust. For $\gamma =\frac 43$,
the equation of state is that of photon gas or a gas of non-interacting
relativistic particles. For $\gamma =2$, the equation of state reduces to
the stiff matter case.

\section{\textbf{Non-Static Spherically Symmetric Solutions}}

To simplify the field equations we solve them for some special cases:

\subsection{\bf{THE CASE} R=R(t), $\lambda =\lambda$(t)}

This class is identical with the well-known Kantowski--Sachs class of
cosmological models [7-9]. We can choose $R=t$ without loss of generality,
use the equation of state given by Eq.(8) and attempt all the possible solutions.

{\textbf{When }$\gamma =1$}

This gives $p=0$ and the EFEs.(3-6) will reduce to
\begin{equation}
\kappa \rho =\frac 1{t^2}+\frac 2te^{-2\nu }(\stackrel{.}{\lambda }+\frac
1{2t}),
\end{equation}
\begin{equation}
0=2t\stackrel{.}{\nu }e^{-2\nu }-1-e^{-2\nu },
\end{equation}

\begin{equation}
0=-e^{-2\nu }[(\stackrel{..}{\lambda }+\stackrel{.}{\lambda }^2-\stackrel{.}{%
\lambda }\stackrel{.}{\nu })t+\stackrel{.}{\lambda }-\stackrel{.}{\nu }].
\end{equation}

Eq.(10) can easily be solved which gives
\[
\,\nu =\frac 12\ln \left| \frac t{c-t}\right|,
\]
where c is an arbitrary constant. For this value of $\nu$, Eq.(9) gives
\[
\,\stackrel{.}{\lambda}=\frac{\rho \kappa t^3-c}{2t(c-t)}.
\]
Substituting the values of $\stackrel{..}{\lambda},
\stackrel{.}{\lambda}^2, \stackrel{.}{\lambda}, \stackrel{.}{\nu}$ in Eq.(11),
we have
\[
\stackrel{.}{\rho}+\frac{3c-4t}{2t(c-t)}\rho =\frac{t^2\kappa }{2(t-c)}
\rho ^2,
\]
Solving this, we get

\begin{equation}
\rho =\left[ \kappa \{(\frac t{c-t})^{\frac 12}-\sin^{-1}(\frac tc)^{\frac 12}\}%
\sqrt{ct^3-t^4}^{}+c_1\sqrt{ct^3-t^4}\right] ^{-1},
\end{equation}
where $c_1$ is an integration constant. By substituting this value of $\rho$
in $\stackrel{.}{\lambda}$ and integrating, the value of $\lambda$ will become
\begin{equation}
\lambda=\ln\left[c_2\{1-(\frac{c-t}{t})^\frac 12 sin^{-1}(\frac tc)^\frac 12+(\frac
{c-t}{t})^\frac 12 \frac{c_1}{\kappa}\}\right],
\end{equation}
where $c_2$ is an integration constant. The energy-momentum conservation relations (7) are also
satisfied by this solution.

The resulting spacetime is
\begin{equation}
ds^2=\frac t{c-t}dt^2-\left[c_2\{1-(\frac{c-t}{t})^\frac 12 sin^{-1}(\frac tc)^\frac 12+(\frac
{c-t}{t})^\frac 12 \frac{c_1}{\kappa}\}\right]^2dr^2-t^2d\Omega ^2
\end{equation}

If we take a special case for which $\lambda = constant$ and $\gamma =2(\rho =p),$
then EFEs (3-6) will give
\[
\nu =\ln \left| \frac{\alpha^{2}t^{2}}{1-\alpha^{2}t^{2}}\right| ^{%
\frac{1}{2}},
\]
where $\alpha$ is an arbitrary constant. The energy density and the pressure
can be evaluated as
\begin{equation}
\rho=p=\frac{1}{\kappa \alpha^2t^4}.
\end{equation}
The corresponding metric will be

\begin{equation}
ds^2=\frac{\alpha^2t^2}{1-\alpha^2t^2}dt^2-dr^2-t^2d\Omega ^2.
\end{equation}

\subsection{\bf{THE CASE} R=R(t), $\lambda =\lambda $ (r,t)}

This class of solution was examined by Korkina and Martinenko [10,11]. In
this case EFEs will be the same as Eqs.(3-6) except that now $\lambda
=\lambda (r,t).$ We solve this system of partial differential equations
using Eq.(8).

{\textbf{When} $\gamma =2$}

In this case $p=\rho$, EFEs (3-6) will give

\begin{equation}
\rho =\frac 1{\kappa t^2}[y(1+\frac{2\stackrel{.}{z}}z t)+1],
\end{equation}

\begin{equation}
p=-\frac 1{\kappa t^2}[\stackrel{.}{y}t+1+y],
\end{equation}

\begin{equation}
p=-\frac 1{\kappa t^2}[y\frac{\stackrel{..}{z}}{z}t^2+\stackrel{.}{y}\frac{
\stackrel{.}{z}}{2z}t^2+y\frac{\stackrel{.}{z}}{z}t+\frac{\stackrel{.}{y}}{2}t],
\end{equation}
where $y=e^{-2\nu (t)},\quad z =e^{\lambda (r,t)}$.
From Eqs.(18) and (19), we have
\begin{equation}
\stackrel{..}{z}+\stackrel{.}{z}[\frac{\stackrel{.}{y}}{2y}%
+\frac 1t]-z[\frac{\stackrel{.}{y}}{2yt}+\frac{y+1}{yt^2}]=0.
\end{equation}
This second order non-linear partial differential equation can be solved
using Herlt's method [5] by choosing
\begin{equation}
y(t)=e^{-2\nu }=\frac 1{n^2-1}+\beta t^{-2(n+1)}, \quad n^2\neq 1
\end{equation}
where $\beta$ is an arbitrary constant. Eq.(20) then becomes
\begin{equation}
\stackrel{..}{z}+A\stackrel{.}{z}-Bz =0,
\end{equation}
where

\begin{equation}
A=\frac{t^{-1}-n(n^2-1)\beta t^{-2n-3}}{1+\beta (n^2-1)t^{-2n-2}}\,,
\end{equation}

\begin{equation}
B=\frac{n^2t^{-2}-n(n^2-1)\beta t^{-2n-4}}{1+\beta (n^2-1)t^{-2n-2}}
\end{equation}
Eq.(22) has the special solution $z_s=t^n.$ Let us substitute
\begin{equation}
z=C(r,t)\,t^n
\end{equation}
into Eq.(22), we obtain

\begin{equation}
\stackrel{..}{C}+[A+2nt^{-1}]\stackrel{.}{C}=0,
\end{equation}
This can easily be solved, and the general solution becomes

\begin{equation}
z(r,t)=t^n\{\beta_1(r)\int_{\beta_2(r)}^t\frac{t^{^{\prime }-n}}{\sqrt{%
t^{^{\prime }2n+2}+\beta (n^2-1)}}dt^{^{\prime }}\},
\end{equation}
where $\beta_1(r)\,$ and $\beta_2(r)$ are arbitrary functions of the variable $r.$
The energy density $\rho $ and pressure $p$ can now be computed easily using Eqs.(17) and (18).

The corresponding metric will become

\begin{equation}
ds^2=[y(t)]^{-1}dt^2-z^2(r,t)^{}dr^2-t^2d\Omega ^2,
\end{equation}
where $y(t)$ and $z \left( r,t\right) \,$ are given by Eqs.(21) and (27) respectively.
It is to be noticed that this solution corresponds to the solution obtained by Herlt [5].

\subsection{THE GENERAL CASE R=R(r,t)}

To solve the general case we take the following two assumptions:
\[
(i)\nu =0,\lambda =\lambda (r,t);\quad (ii)\nu =\nu
(r,t),\lambda =0.
\]

\textbf{(i) When }$\nu =0,\lambda =\lambda (r,t)$

EFE (6) implies that $\lambda =\ln \left| fR^{^{\prime }}\right|,$
where $f(r)$ is an arbitrary function of $r$ and $R$ is an arbitrary function of
coordinates $r$ and $t.$ Also $R^{\prime }\neq 0,$ which implies that $R\neq R(t).$

Here arises two cases i.e. either $R=R(r)$ or $R=R(r,t).$
For $R=R(r),$ we can have $\rho$ and $p$ by replacing $\lambda$ in
Eqs.(3) and (4) respectively.
\[
\rho =\frac 1\kappa (\frac 1{R^2}-\frac 1{f^2R^2}+\frac{2f^{^{\prime }}}{%
R^{}R^{^{\prime }}f^3}),
\]
\[
p=\frac 1\kappa (-\frac 1{R^2}+\frac 1{f^2R^2}).
\]
But Eq.(5) gives
\[
p=-\frac{f^{^{\prime }}}{\kappa RR^{^{\prime }}f^3}.
\]
By comparing the two values of $p,$ we obtain
\[
R=\frac{\sqrt{f^2-1}}{lf},
\]
where $l$ is an integration constant. The resulting metric becomes
\begin{equation}
ds^2=dt^2-\frac{f'^2}{l^2f^2(f^2-1)}dr^2-\frac{f^2-1}{l^2f^2}d\Omega^2.
\end{equation}
This turns out a class of spherically symmetric static spacetimes. For $f=\frac{1}{\sqrt{1-r^2}}$ and
$l=1$, it reduces to Einstein metric. From the above equations we obtain $\rho=\frac{3l^2}{\kappa},\quad
p=-\frac{l^2}{\kappa}$ which implies that $\rho +3p=0$.

\textbf{(ii)When} $\nu =\nu (r,t), \lambda =0$,
EFE (6) will reduce to $\nu =\ln \left| g\stackrel{.}{R}\right|,$
where g(t) is an arbitrary function of \thinspace t and R is an arbitrary
function of coordinates r and t. Also $\stackrel{.}{R}\neq 0,$ which
implies that $R\neq R(r).$ This gives that either $R=R(t),R=R(r,t).$ For $R=R(t),$
we have the spacetime similar to Eq.(16). The quantities $\rho$ and p also turn out to be the same.

For $R=R(r,t)$, the solutions need to be investigated.

\section{\bf{Kinematic of the Velocity Field}}

The spherically symmetric solutions can be classified according to
their kinematical properties [6]. The rotation is given by

\begin{equation}
\omega_{ab}=u_{[a;b]}+\stackrel{.}{u_{[a}}u_{b]}.
\end{equation}
Tha acceleration can be found by

\begin{equation}
\stackrel{.}{u_a}=u_{a;b}u^b.
\end{equation}
For the expansion we have

\begin{equation}
\Theta=u^a_{;a}.
\end{equation}
The components of the shear-tensor are given by

\begin{equation}
\sigma_{ab}=u_{(a;b)}+\stackrel{.}{u}_{(a}u_{b)}-\frac 13 \Theta h_{ab},
\end{equation}
where $h_{ab}=g_{ab}-u_a u_b$ is the projection operator. The square brackets denote antisymmetrization
and the round brackets indicate symmetrization.
The shear invariant is given as $\sigma_{ab}\sigma^{ab}$.

Now we find all the above quantities for the solutions obtained.
The rotation and the acceleration are zero for all the solutions.
The expansion, for the first solution, is

\begin{equation}
\Theta=(\frac{c-t)}{t})^{\frac 12}[\frac{\rho \kappa t^3-c}{2t(c-t)}+\frac 2t]
\end{equation}
The components of the shear-tensor are given by

\begin{equation}
\sigma_{11}=\frac 23(\frac{c-t)}{t})^{\frac 12}[\frac 1t-\frac{\rho \kappa t^3-c}{2t(c-t)}]
\left[c_2\{1-(\frac{c-t}{t})^\frac 12 sin^{-1}(\frac tc)^\frac 12+(\frac{c-t}{t})^\frac 12
\frac{c_1}{\kappa}\}\right]^2,
\end{equation}

\begin{equation}
\sigma_{22}=\frac 13 t^2(\frac{c-t)}{t})^{\frac 12}[\frac{\rho \kappa t^3-c}{2t(c-t)}-\frac 1t],
\end{equation}

\begin{equation}
\sigma_{33}=\sin^2\theta \sigma_{22}.
\end{equation}
For the second solution, the expansion factor is

\begin{equation}
\Theta=\frac{2(1-\alpha^2t^2)^{\frac 12}}{\alpha t^2}
\end{equation}
The components of the shear-tensor are given by

\begin{equation}
\sigma_{11}=\frac{2(1-\alpha^2t^2)^{\frac 12}}{3\alpha t^2},
\end{equation}
\begin{equation}
\sigma_{22}=-\frac{(1-\alpha^2t^2)^{\frac 12}}{3\alpha},
\end{equation}
\begin{equation}
\sigma_{33}=\sin^2\theta \sigma_{22}.
\end{equation}
For the third solution, we have

\begin{equation}
\Theta=(y)^{\frac 12}(\frac {\stackrel{.}{z}}{2z}+\frac 2t)
\end{equation}
The components of the shear-tensor are given by
\begin{equation}
\sigma_{11}=\frac 23(\frac 1t-\frac {\stackrel{.}{z}}{2z})z\sqrt y,
\end{equation}
\begin{equation}
\sigma_{22}=\frac 13 t^2(\frac {\stackrel{.}{z}}{2z}-\frac 1t)\sqrt y,
\end{equation}

\begin{equation}
\sigma_{33}=\sin^2\theta \sigma_{22}.
\end{equation}
The shear invariant turns out to be 3 for all the solutions.

The rate of change of expansion with respect to proper time is given by Raychaudhuri's
equation [12]
\begin{equation}
\frac{d\Theta}{d\tau}=-\frac 13 \Theta^2-\sigma_{ab}\sigma^{ab}+\omega_{ab}u^au^b-R_{ab}u^au^b
\end{equation}
We evaluate it only for the second solution as it is simple to understand. For this solution it becomes
\begin{equation}
\frac{d\Theta}{d\tau}=-\frac{2}{\alpha^2}(\frac{1+\alpha^2t^2}{t^4})
\end{equation}
\section{\textbf{SUMMARY}}

This section presents a brief summary of the results obtained in the
previous section. We shall discuss these results and will comment on
possible future directions of development.

We have been able to solve partially the EFEs for the classes of non-static
spherically symmetric spacetimes using different equations of state for
perfect fluid. These have been classified into three categories. The summary
of each case is given below:

\textbf{1.} $\mathbf{R=R(t),}$\textbf{\ }$\mathbf{\lambda =\lambda (t)}$

In this case, the dust solution gives

\begin{equation}
ds^2=\frac t{c-t}dt^2-\left[c_2\{1-(\frac{c-t}{t})^\frac 12 sin^{-1}(\frac tc)^\frac 12+(\frac
{c-t}{t})^\frac 12 \frac{c_1}{\kappa}\}\right]^2dr^2-t^2d\Omega ^2
\end{equation}

The stiff matter solution becomes

\begin{equation}
ds^2=\frac{\alpha^2t^2}{1-\alpha^2t^2}dt^2-dr^2-t^2d\Omega ^2.
\end{equation}

\textbf{2.} $\mathbf{R=R(t),}$\textbf{\ }$\mathbf{\lambda =\lambda (r,t)}$

The stiff matter solution gives

\begin{equation}
ds^2=[y(t)]^{-1}dt^2-z^2(r,t)dr^2-t^2d\Omega ^2,
\end{equation}
where y and z are already given.

\textbf{3.} $\mathbf{R=R(r,t)}$

In this case we obtain two solutions. The first solution, infact, turns out a class of spherically
symmetric static spacetimes. The other (stiff matter) solution is exactly the same as Eq.(49).

The non-static spherically symmetric solutions with equations of state split
up into three classes of solutions. In the first case we obtain two different
solutions. One of these two becomes the dust solution and the other becomes a
stiff matter solution. The pressure and the energy density are positive
every where. In case two we have a stiff matter solution only. However,
it is much difficult to comprehend it. In the 3rd case we obtain two solutions. The first
solution becomes a class of spherically symmetric static spacetimes depending upon the
arbitrary function f. If we choose $f=\frac{1}{\sqrt{1-r^2}}$, it reduces to Einstein
spacetime. The energy density is positive everywhere while the pressure is negative. The
other (stiff matter) coincides with one of the solutions in the first case. It is
interesting to note that all the non-static solutions which have been found contain shear
which are a very few solutions in the literature.

Finally, we discuss the behaviour of the rate of change of expansion for one solution
given by Eq.(49). We see from Eq.(47) that the rate will never be positive but this will
always be negative. As the time t tends to zero, the rate approaches to $-\infty$ and as t
goes to $\infty$, the rate tends to zero. This shows that the spacetime is contracting or
collapsing and the flux gets focussed along the proper time. Similarly, the other
solutions can be discussed.

We have tried to obtain non-static spherically symmetric solutions
for some particular classes and partial solutions have been attempted in these three
classes. Thus a total of three solutions have been obtained. To obtain new
solutions one has to solve the remaining cases of these classes. Then we can
attempt the general solution of non-static spherically symmetric solutions.
\newpage\

\begin{description}

\item {\bf Acknowledgment}

One of the authors (MS) would like to thank Prof. Chul H. Lee for the hospitality at the Department of
Physics and the Korea Science and Engineering Foundation (KOSEF) for the postdoc fellowship at
Hanyang University Seoul, KOREA, where some of this work was completed.

\end{description}

\vspace{2cm}

{\bf \large References}

\begin{description}

\item{[1]}  Knutsen, H.: {\it Gen. Rel. Grav.} {\bf 24}(1992)1297.

\item{[2]}  Knutsen, H.: {\it Class. Quant. Grav.} {\bf 11}(1995)2817.

\item{[3]}  Kitamura, S.: {\it Class. Quant. Grav.} {\bf 11}(1994)195.

\item{[4]}  Kitamura, S.: {\it Class. Quant. Grav.} {\bf 12}(1995)827.

\item{[5]}  Herlt, Eduard: {\it Gen. Rel. Grav.} {\bf 28}(1996)919.

\item{[6]}  Kramer, D. et al: {\it Exact Solutions of Einstein's Field
Equations }(Cambridge University Press 1980).

\item{[7]}  Kompaneets, A. and Chernnow, A.S.: {\it ZhETF} {\bf 47}(1964)1939,
{\it [Sov. Phys. JETP} {\bf 20}(1965)1303].

\item{[8]}  Kantowski, R. and Sachs, R. K.: {\it J. Math. Phys}. {\bf 7}(1966)443.

\item{[9]}  Krasinki: {\it A Physics in An Inhomogeneous Universe}
(Cambridge University Press 1996).

\item{[10]}  Korkina, M. P., Martinenko, V.G.: {\it Ukr. Fiz. Zh.}
{\bf 20}(1975)626.

\item{[11]}  Korkina, M. P., Martinenko, V.G.: {\it Ukr. Fiz. Zh.}
{\bf 20}(1975)2044.

\item{[12]}  Wald, R.M.: {\it General Relativity} (University of Chichago Press Chichago,
1984).

\end{description}

\end{document}